\documentclass[aps,prl,twocolumn,showpacs]{revtex4}
\usepackage[dvips]{graphicx}
\usepackage{amsmath}
\usepackage[latin1]{inputenc}
\usepackage[english,frenchb]{babel}
\usepackage{ulem}

\begin{document}

\title{Study of light-assisted collisions between a few cold atoms in a microscopic dipole trap}

\author{A. Fuhrmanek, R. Bourgain, Y.R.P. Sortais and A. Browaeys}

\affiliation{Laboratoire Charles Fabry, Institut d'Optique, CNRS, Univ Paris Sud, 2 Avenue Augustin Fresnel,
91127 PALAISEAU cedex, France}

\date{\today}

\begin{abstract}

We study light-assisted collisions in an ensemble containing a small number ($\sim 3$) of cold $^{87}$Rb atoms trapped in a microscopic dipole trap.
Using our ability to operate with one atom exactly in the trap, we measure the one-body heating rate associated to a near-resonant laser excitation, and we use this measurement to extract the two-body loss rate associated to light-assisted collisions when a few atoms are present in the trap. Our measurements indicate that the two-body loss rate can reach surprisingly large values $\beta>10^{-8}$cm$^{3}$s$^{-1}$ and varies rapidly with the trap depth and the parameters of the excitation light.
\end{abstract}


\maketitle

\section{I. Introduction}

Extensive experimental and theoretical studies have been devoted in the last decades to light-assisted collisions, using cold atoms held in a magneto-optical trap~\cite{WeinerRMP}, including at the few atom level~\cite{Ueberholz2000}, or in large optical dipole traps~\cite{Kuppens2000,Kulatunga2010}. In small dipole traps with size comparable to the wavelength of the light, light-assisted collisions are used to prepare or probe mesoscopic atomic ensembles, opening new avenues in condensed matter physics and quantum information processing. For instance, they are at the heart of the preparation of individual atoms in microscopic optical
dipole traps~\cite{Schlosser2001,Grunzweig2010,Birkl2010}, standing waves~\cite{Forster2006} or three dimensional optical lattices ~\cite{Weiss2007}. They are also at the origin of sub-poissonian atom number distributions in a mesoscopic atomic ensemble~\cite{Sortais2012}. Finally, they constitute an important tool to understand quantum phases, as demonstrated recently with atoms in optical lattices~\cite{Bakr2010,Bloch2010}. While conceptually simple, the theoretical
description of light-induced collisions is known to be cumbersome due to the complex interplay between atomic multi-level structure and atom-light coupling.
The situation is even worse when considering tightly-confined atomic ensembles where the trapping potential acts on the same length scale as the interaction between the atoms. As a consequence no theoretical prediction for the loss rates and their dependency on parameters such as the atomic density or the light parameters is available to date for this system. The absence of reported measurements makes the situation even more dramatic.

In this paper, we report on an experimental study of light-assisted collisions between cold atoms that are tightly confined in a microscopic dipole trap. To allow for future theoretical modeling of our data, we implemented as closely as possible the gedanken experiment where merely two atoms in the ground state (here, $^{87}$Rb in state $5S_{1/2}~F=2$) collide in the presence of a nearly-resonant laser field. In our case, the loading of the dipole trap is non deterministic~\cite{Sortais2012} and we operate with a typical average atom number of $\sim 3$. We then illuminate the trapped atoms with a pulse of near-resonant light with known frequency and intensity, in order to trigger losses. The near-resonant light has two effects: it heats the atoms individually out of the trap and it induces two-body losses, which we wish to study. To separate the two contributions, we proceed in two steps. First, we use our ability to operate with exactly one atom to measure the one-body heating. Second, we operate with $\sim 3$ trapped $^{87}$Rb atoms and use the result of the single-atom
measurement to extract the two-body loss rate. To extract this rate we develop a Monte-Carlo simulation that we compare to the data. Our measurement indicates
light-assisted collision rates that can reach remarkably large values ($\sim 10^{-8}{\rm cm}^3.{\rm s}^{-1}$), well above measured data found in the literature (by $1-2$ orders of magnitude) for atoms held in magneto-optical traps (for a review, see~\cite{WeinerRMP}) or in larger dipole traps~\cite{Kuppens2000,Kulatunga2010}. Our maximal light-assisted collision rates are surprisingly close to the semi-classical Langevin limit.

\section{II. Principle of the experiment}
\begin{figure}
\includegraphics[width=8cm]{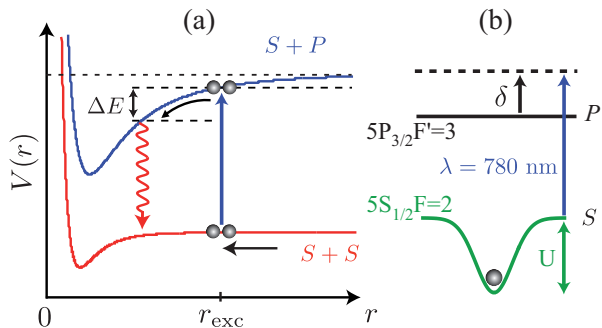}
\caption{(a) Light-assisted atom loss by radiative escape. Atoms in the $S$ state interact through the Van der Waals potential $V(r)=-C_6/r^6$ while atoms in the $S$ and $P$ states interact through the dipole-dipole attractive potential $V(r)=-C_3/r^3$. (b) Levels involved in the experiment. The excitation laser detuning $\delta$ is measured with respect to the free space transition.} \label{Fig:figure1}
\end{figure}
In our apparatus we operate an optical dipole trap at $850$~nm with micrometer size (waist $1~\mu$m)~\cite{Sortais2007}, which we load with cold atoms from a magneto-optical trap (MOT). Atoms enter the microscopic trap randomly, are trapped thanks to the cooling effect of the MOT beams, and are expelled from the trap due to one- or two-body processes. Depending on the local density of the MOT cloud around the dipole trap, we control the number of trapped atoms in steady state from one atom exactly ($N_0=1$) to a few atoms on average ($\langle N_0\rangle\simeq 3$)~\cite{Fuhrmanek2010}.

To study the light-assisted collisions, we switch off the MOT beams and then send the pulse of excitation light on the trapped atoms, initially prepared in the $5S_{1/2},F = 1$ level. The excitation light consists of repumping light that transfers the atoms to the ($5S_{1/2},F = 2$) level, labeled $S$ in
Fig.~\ref{Fig:figure1}~\footnote{The repumping light is kept on resonance with the $(5S_{1/2},F = 1)\rightarrow (5P_{3/2},F' = 2)$ transition in free space and has a saturation of $\sim 20$.}, superimposed with light nearly resonant with the $(5S_{1/2},F = 2) \rightarrow (5P_{3/2},F' = 3)$ light-shifted transition, which excites the atoms into the $(5P_{3/2},F' = 3)$ level, labeled $P$ (see Fig.~\ref{Fig:figure1}b). The excitation light consists of a pair of counter-propagating
laser beams with orthogonal circular polarizations. For this experiment, we do not control the orientation of the magnetic field, which has a magnitude smaller than $0.2$~G.

During the laser excitation, two atoms form a loosely bound pair with one atom in the $S$ state and the other in the $P$ state and interact through the long-range  dipole-dipole attractive potential $V(r)=-C_3/r^3$ (here, $r$ is the inter-atomic distance, $C_3=3\hbar\Gamma/4k^3$, $\Gamma/2\pi= 6$~MHz is the linewidth of the $P$ state, $\lambda=2\pi/k=780$\,nm is the wavelength of the $S\rightarrow P$ transition), as represented in Fig.~\ref{Fig:figure1}a. If the kinetic energy acquired by an atom pair before it radiates back to the gound state exceeds the optical dipole trap depth $U$, it escapes the trap, thus leading to the loss of two atoms. This interaction-induced loss mechanism, known as radiative escape\,\cite{GallagherPritchard89}~\footnote{We neglect here fine structure-changing collisions that occur at much shorter interatomic distances~\cite{WeinerRMP}.}, co-exists with the standard one-body loss mechanism associated to the cycles of absorption and spontaneous emission of photons by individual atoms in the trap, which heat them out of the trap. Whether one or the other mechanism is dominant depends on the parameters of the experiment, namely the trap depth $U$, the saturation $s=I/I_{\rm sat}$ ($I$ is the laser intensity and $I_{\rm sat}=1.6$~mW/${\rm cm}^2$) and the frequency detuning $\delta$ of the excitation light with respect to the single atom transition in free space (see Fig.~\ref{Fig:figure1}b).

For a given set of parameters we measure the number of trapped atoms that remain in the trap after the pulse of light has been sent. The number of atoms is measured by accumulating their fluorescence at $780$~nm on an intensified CCD camera and comparing it to the calibrated fluorescence of a single atom~\cite{Fuhrmanek2010NJP,Fuhrmanek2010}. By varying the duration $t$ of the pulse we obtain atom loss curves, from which we extract the one- and two-body loss rates as explained below.

\section{III. Light-induced losses of single atoms.}
\begin{figure}
\includegraphics[width=8cm]{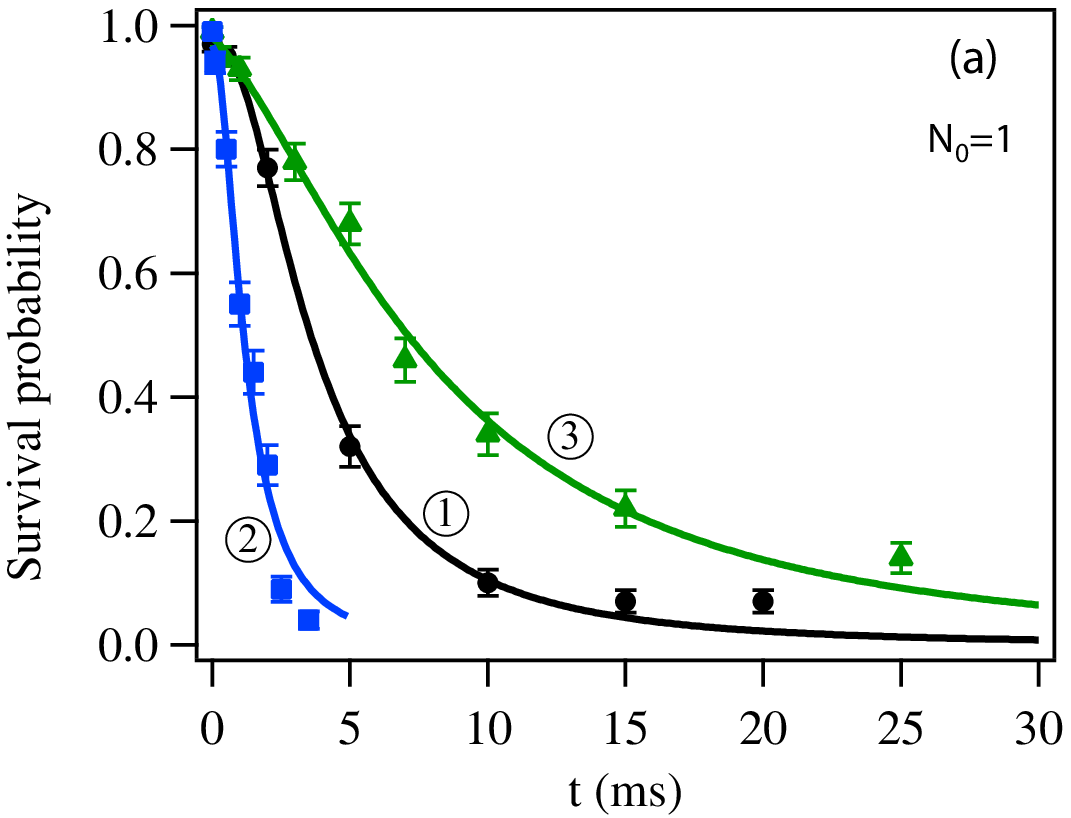}
\includegraphics[width=8cm]{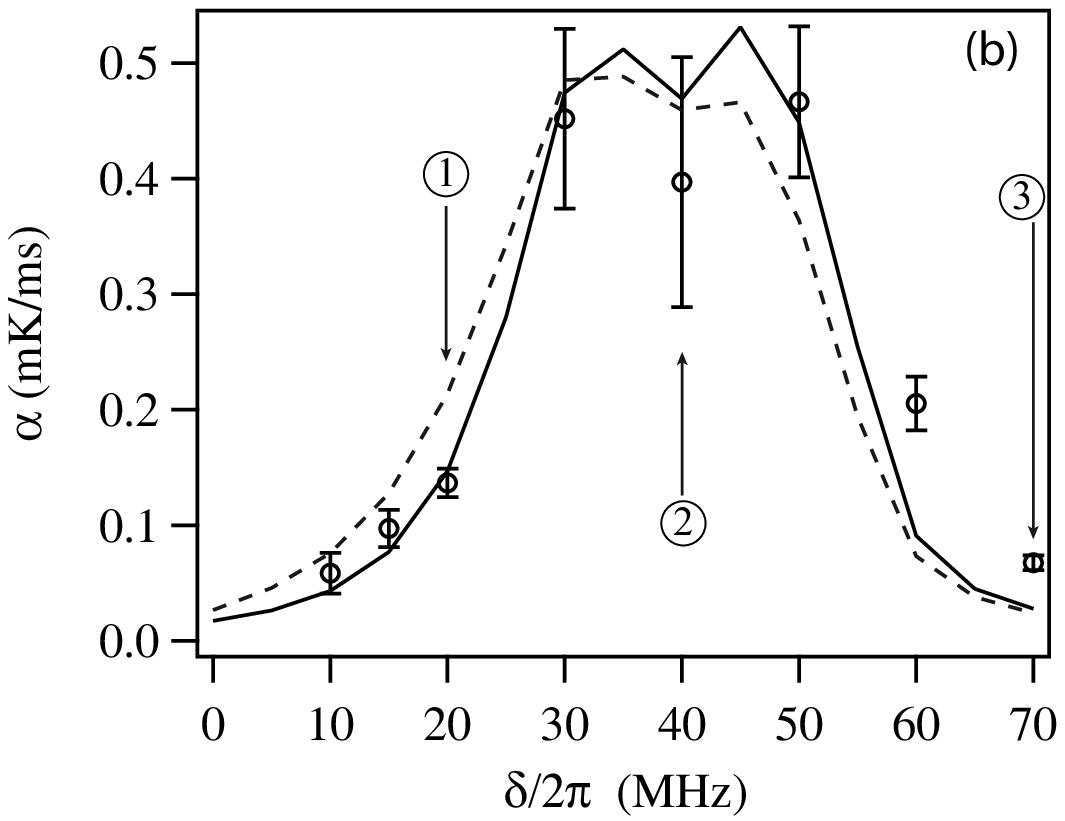}
\caption{(a) Survival probability $P_{1}(t)$ of a single atom in the dipole trap, measured for various detunings of the excitation light $\delta/2\pi =\{20;40;70\}~$MHz (circles, squares and triangles, respectively). The trap depth is $U =h\times 36$~MHz$=k_\textrm{B}\times 1.8$~mK and the saturation parameter is $s=0.5$. Solid lines show fits of the data to a heating model (see text). The initial temperature of the single atom is $T_0=200~\mu$K. Error bars are statistical. (b) Heating rate $\alpha$ (circles) deduced from the fits of the survival probability, versus $\delta/2\pi$ (error bars are from the fit); the solid and dashed lines are solutions of the rate equation model described in the text, with no free parameter, for $T_0=200~\mu$K and $T_0=300~\mu$K respectively. Data labeled $1$, $2$ and $3$ are extracted from the loss curves shown in (a).}\label{Fig:DecayCurve_Single}
\end{figure}

To extract the one-body loss rate, we perform the loss experiment described above with one atom exactly in the trap ($N_0=1$). To do this, we adjust the loading rate of the trap to operate in the collisional blockade regime~\cite{Schlosser2002,Sortais2007} and trigger the loss experiment on the presence of a single atom in the trap. All parameters (trap depth, excitation light parameters) are otherwise unchanged with respect to the case $\langle N_0\rangle=3$ explored later in this paper. We obtain the survival probability of a single atom after the excitation process by repeating the experiment $200$ times and measuring each time the presence or the absence of the atom in the trap after the experiment. Figure~\ref{Fig:DecayCurve_Single}a shows examples of loss curves that illustrate the effect of the excitation light. For comparison, the lifetime of the atom in the trap is $24$~s in the absence of the excitation light and is limited by the residual background gas collisions.

The effect of the excitation light is to heat the atom out of the trap as the duration of the excitation increases. This effect is quantitatively well explained by assuming that the temperature of the atom varies in time as $T(t) = T_0+ \alpha t$ since the energy of the atom increases linearly with each absorption and spontaneous emission cycle. Here, $T_0$ is the temperature at the beginning of the excitation pulse and $\alpha$ is the heating rate. Assuming a harmonic trap and a Boltzmann energy distribution~\cite{Tuchendler2008}, the probability $P_1(t)$ for a single atom to remain in the trap with depth $U$ at a temperature $T(t)$ is given by
\begin{equation} \label{equationP1}
P_{1}(t) = 1-\left[ 1 + \eta(t) + \frac{1}{2}\eta (t)^2
\right]\exp\left(-\eta(t)\right)
\end{equation}
where $\eta(t) = U/k_{\rm B} T(t)$ ($k_{\rm B}$ is the Boltzmann constant). The temperature $T_0$ being measured independently by a
release-and-recapture technique~\cite{Tuchendler2008}, we fit the data to the function $P_1(t)$ with $\alpha$ being the only free parameter. The result obtained for $\alpha$ is shown in  Fig.~\ref{Fig:DecayCurve_Single}b. As expected, it reaches a maximum when the excitation light is nearly-resonant with the light-shifted $S\rightarrow P$ transition, i.e. $\delta=U/\hbar$ ($h=2\pi\hbar$ is the Planck constant).

The position and the shape of this resonance is confirmed by calculating the heating rate $\alpha = 2 E_{\rm r}. R/k_{\rm B}$, where  $R$ is the photon scattering rate and $E_{\rm r}$ is the recoil energy. We use a rate equation model to calculate the populations of the various Zeeman sub-levels from which we deduce the scattering rate $R$. We take into account the finite initial temperature of the atom $T_0=200~\mu$K and its linear increase in time, which lead to random positions of the atom in the trap, and therefore to different light-shifts for the various Zeeman states. As seen in Fig.~\ref{Fig:DecayCurve_Single}b, this model reproduces the broadening of the resonance with respect to the natural line width $\Gamma/2\pi=6$~MHz. In the following, we will use it to extrapolate the heating rate $\alpha$ to other values of $T_0$, as $T_0$ varies when operating with more than one atom initially. For instance, the dashed curve in Fig.~\ref{Fig:DecayCurve_Single}b shows how the resonance in $\alpha$ is shifted towards the low values of $\delta$ when $T_0$ increases from $T_0=200~\mu$K (as measured when $N_0=1$) to $T_0=300~\mu$K (when $\langle N_0\rangle=3$). Qualitatively, this shift corresponds to the shift of the Boltzmann energy distribution towards the shallower parts of the trap.

\section{IV. Light-induced losses in the multi-atom case.}
\subsection{A. Experimental observations}
\begin{figure}
\includegraphics[width=8cm]{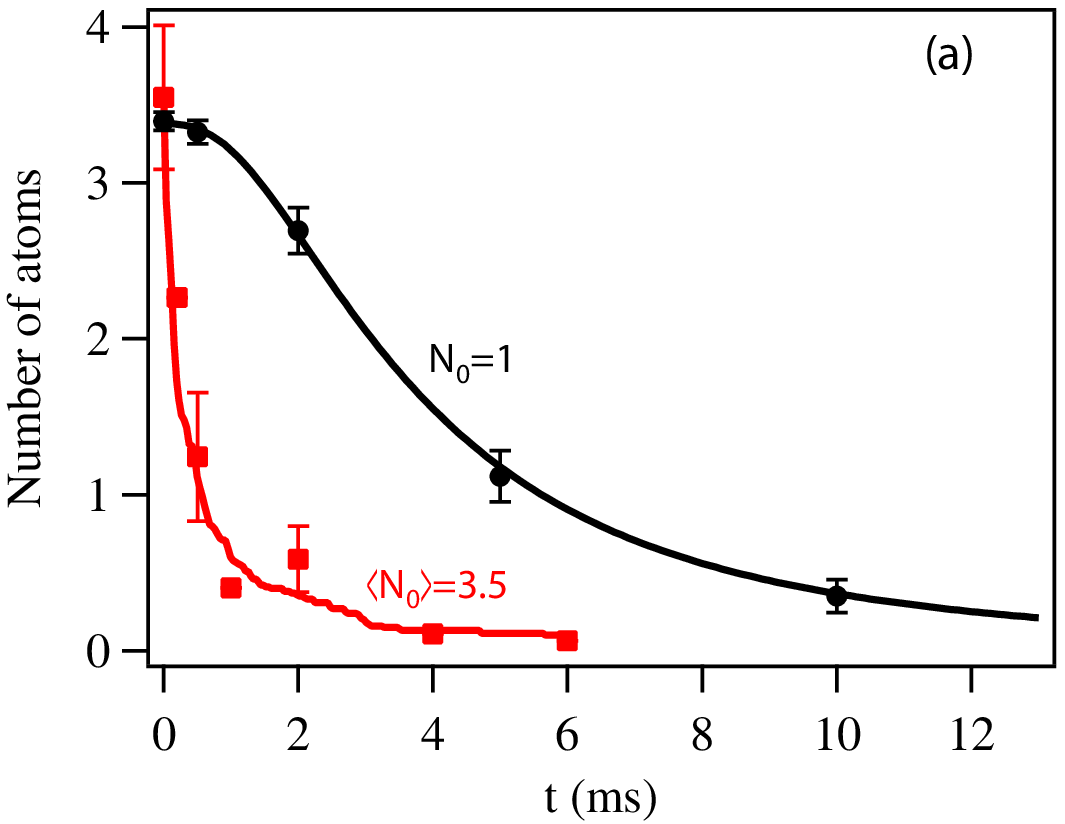}
\includegraphics[width=8cm]{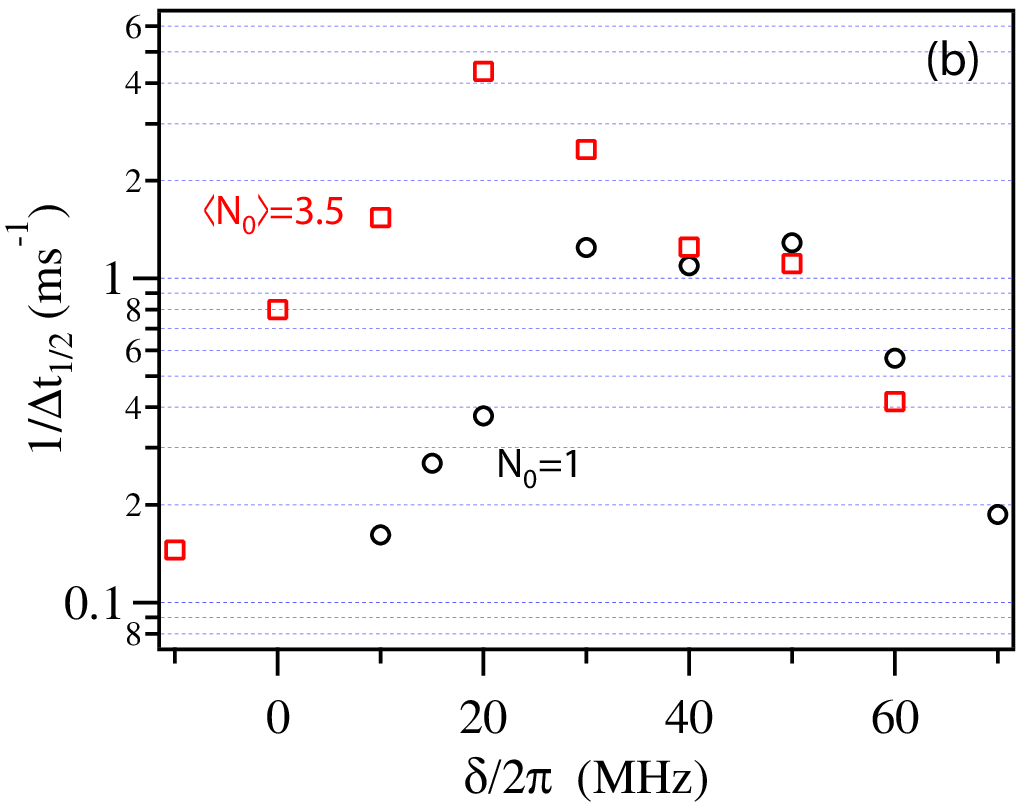}
\caption{(a) Average number of atoms $\langle N(t)\rangle$ remaining in the dipole trap after excitation by a light pulse with duration $t$. Squares: the initial number of atoms in the trap is $\langle N_0\rangle=3.5$; the solid line is a fit to a Monte-Carlo simulation (see text), with a heating rate $\alpha=0.21$~mK/ms and a two-body loss rate $\beta'=1.2\,({\rm at.ms})^{-1}$. Circles: single-atom measurements rescaled to $N_0=3.5$; the solid line is a fit to the radiative heating model. Error bars are statistical. Parameters of the experiments: $U/h=36$~MHz, $\delta/2\pi=20$~MHz, $s=0.5$. (b) Inverse of the half-lifetime in the cases $\langle N_0\rangle=3.5$ (squares) and $N_0=1$ (circles) versus the detuning of the excitation light.}\label{Fig:DecayCurve_Multi}
\end{figure}

We now turn to the case where a few atoms are loaded in the trap. Starting with $\langle N_0\rangle\simeq 3$ atoms, we measure the average number of atoms that remain in the trap after the excitation pulse has been sent~\footnote{The uncertainty on the mean number of initially trapped atoms is $\pm0.4$ due to day to day fluctuations in the loading rate.}. Again the measurements are performed by averaging over several hundreds of experiments. In the absence of excitation light the number of atoms remains constant on time scales large with respect to the pulse duration. In the presence of excitation light we observe losses that can be much faster than in the single atom case, depending on the trap depth and the excitation light parameters. For example, Fig.~\ref{Fig:DecayCurve_Multi}a compares loss curves taken for $(N_0=1,T_0=200~\mu$K$)$ and $(\langle N_0\rangle=3.5, T_0=300~\mu$K$)$ initially, all other parameters being the same ($s=0.5$, $\delta/2\pi=20$\,MHz, $U/h=36$\,MHz). For this set of parameters and $\langle N_0\rangle=3.5$, the number of atoms drops by a factor $2$ in only $0.25$~ms,
at least one order of magnitude faster than in the single atom case. This rapid decrease is incompatible with the radiative heating rate $\alpha$ measured in the single atom case, taking into account the increase in $T_0$ when we operate with a few atoms (see Sec. III and Fig.~\ref{Fig:DecayCurve_Single}b). More generally, we observe this phenomenon for small values of the detuning, typically $\delta\lesssim U/\hbar$ (see Fig.~\ref{Fig:DecayCurve_Multi}b). While a model
involving only the radiative heating process does not reproduce the data, adding two-body losses to the model does, as shown in Fig.~\ref{Fig:DecayCurve_Multi}a. We thus attribute the observed excess losses in this regime to the leading two-body light-induced collisions, neglecting  higher-body collisional processes.
By contrast, for $\delta\gtrsim U/\hbar$, loss curves overlap well in the few- and single-atom cases, indicating that the radiative heating process is identical in both cases and is the dominant loss mechanism. In order to reveal the range of detunings where two-body losses dominate over one-body heating, we have represented in Fig.~\ref{Fig:DecayCurve_Multi}b the inverse of the half-lifetime of the survival probability for both the single and the few-atom cases~\footnote{The data shown in Fig.~\ref{Fig:DecayCurve_Multi}b in the single atom case have been properly extrapolated from the data of Fig.~\ref{Fig:DecayCurve_Single}b to $T_0=300~\mu$K, thus showing the real contribution of radiative heating in the few-atom case.}: for $\delta/2\pi\ge40$~MHz $\approx U/h$, the two curves are nearly identical, indicating that the heating is dominant, while the two-body light-assisted losses
dominate for smaller detunings.

\subsection{B. Model including radiative heating and two-body losses}
To extract the contribution of the two-body loss processes from our loss curves, we developed a Monte Carlo simulation to find  the time-dependent number of atoms $N(t)$. This approach is particularly appropriate in our situation as the competing radiative heating process leads to a time-dependent one-body loss rate and because we need to take into account the discreteness of the small atom number. We describe here the main lines of our simulation.

At each time step $dt$ we evaluate the infinitesimal probability for a two-body loss event to have occurred within the N-atom ensemble between times $t$ and $t+dt$, and we compare it to the probability of a one-body event to have occurred due to radiative heating during the same time interval $dt$. The first is denoted $dq_{\textrm{2-body}}$ and is related to the number of atom pairs at time $t$ and the two-body loss constant $\beta'$ through~\cite{Kuppens2000} $$dq_{\textrm{2-body}}=\beta' N(t)(N(t)-1)d t/2.$$ The second is denoted $dq_{\textrm{1-body}}$ and is related to the number of atoms $N(t)$ and to the instantaneous one-body loss rate $\gamma(t)$ through $$dq_{\textrm{1-body}}=\gamma(t)N(t)dt.$$ Here, $\gamma(t)=-\dot{P}_1(t)/P_1(t)$, as obtained by a Taylor expansion of $P_1(t+dt)$, where $P_1(t)$ is given by Eq.~\ref{equationP1}. In practice, we calculate $P_1(t)$ by using the heating rate $\alpha$ measured in the single-atom regime, corrected by the temperature $T_0$ of the N-atom ensemble, which we measure independently by a time-of-flight method. At each time step of the simulation, three channels are possible: (i) no loss occurs during $dt$: the probability associated to this channel is $(1-dq_{\textrm{1-body}})(1-dq_{\textrm{2-body}})$; (ii) a one-body loss takes place and the atom number decreases by one: the associated probability is $dq_{\textrm{1-body}}(1-dq_{\textrm{2-body}})$;  (iii) a two-body loss occurs and the atom number decreases by two:  the associated probability is $dq_{\textrm{2-body}}(1-dq_{\textrm{1-body}})$.  We pick up randomly one out of these three channels according to their  associated probabilities, calculate the number of atoms at time $t +dt$,  and then proceed to the next time step. By averaging over the initial atom  number distribution $\sim 200$ times we obtain a loss curve that  simulates the actual measurements described above  (see e.g. Fig.~\ref{Fig:DecayCurve_Multi}a).
\begin{figure}
\includegraphics[width=8cm]{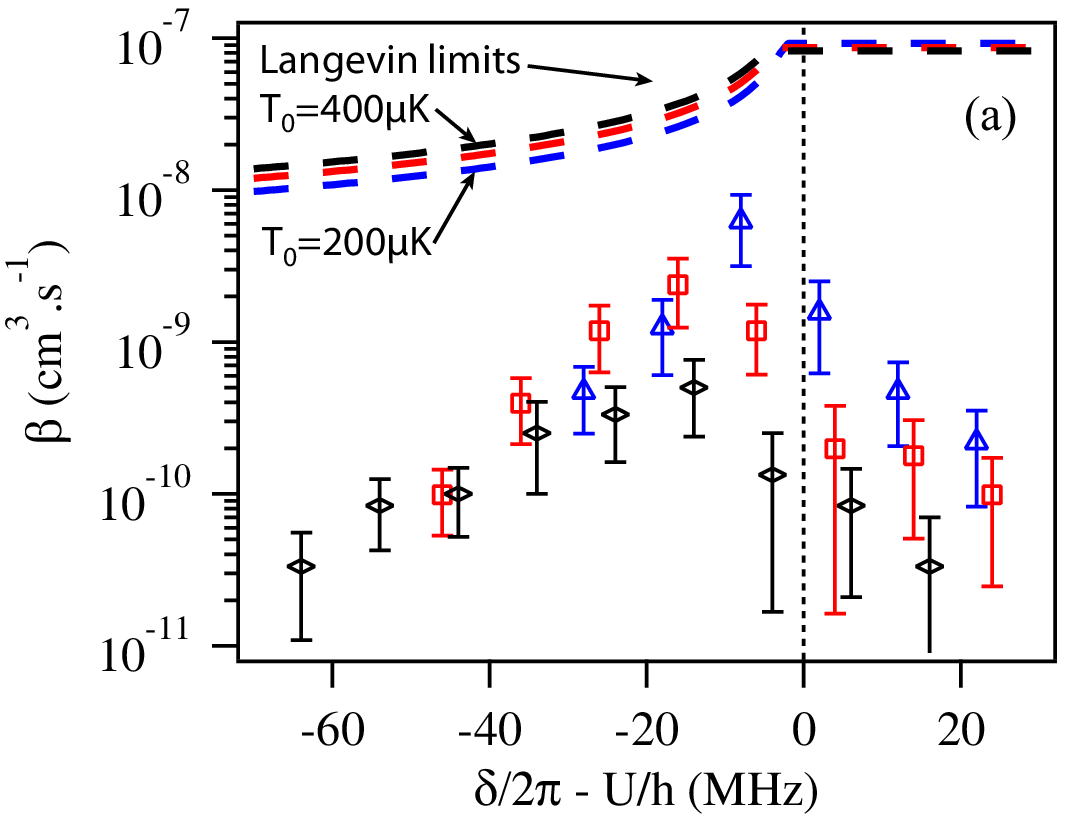}
\includegraphics[width=8cm]{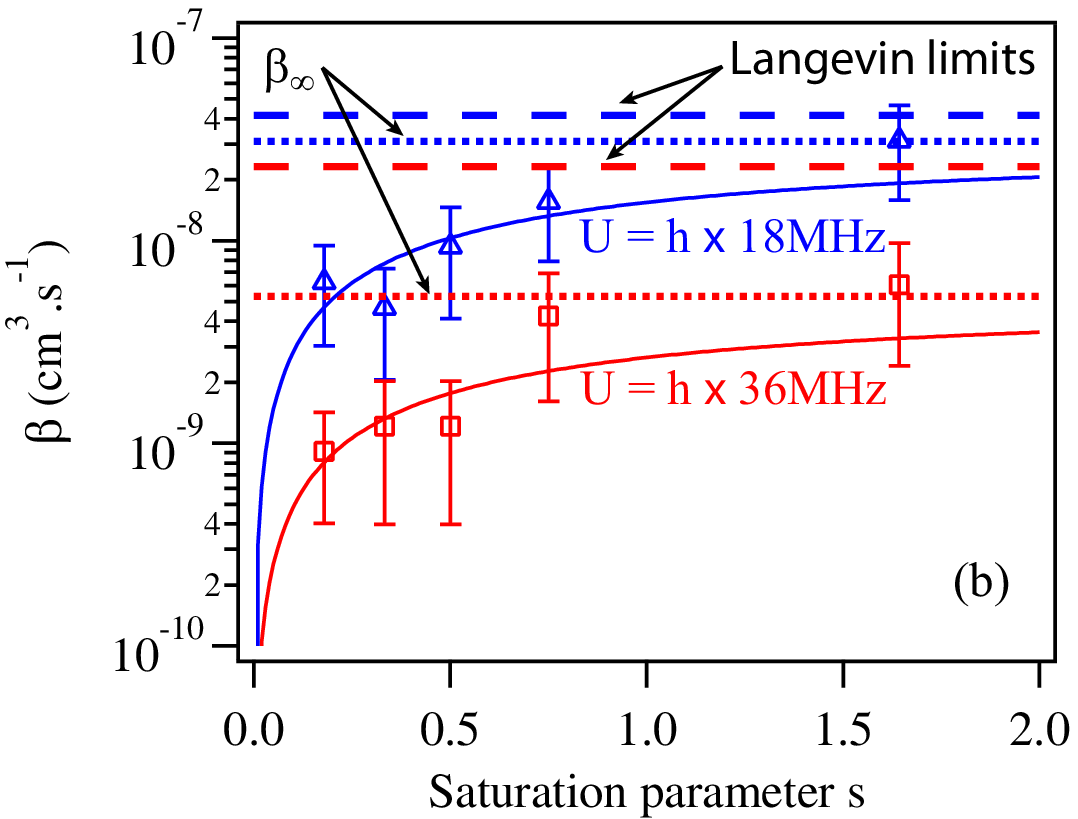}
\caption{Two-body loss rate versus (a) the light-shifted detuning $\delta/2\pi-U/h$, and (b) the saturation parameter $s$ of the excitation light. Triangles, squares and lozenges correspond respectively to trap depths $U/h=(18;36;54)$~MHz, initial peak atomic densities $n_0=(2.7;4.3;5.1)\times 10^{12}$~cm$^{-3}$ and initial temperatures $T_0=(200;300;400)~\mu$K. The average initial atom number is $\langle N_0\rangle\simeq3$ for all the data. Error bars are from the fits. The dashed lines are the Langevin limits $\beta_L$ associated to each set of parameters $(U, T_0)$ (see text). In (a), the saturation parameter is $s= 0.5$. In (b), the detuning of the excitation is $\delta/2\pi=10$~MHz. The solid lines are fits of the data to a $\beta_{\infty}~s/(1+s)$ model (with $\beta_{\infty}$ the only free parameter). The dotted lines indicate the asymptotical value $\beta_{\infty}$ in each case.}\label{Fig:SummaryBeta}
\end{figure}

\subsection{C. Light-assisted two-body loss rates}
The Monte Carlo simulation described in the previous section yields values of $\beta'$ that range from $0.02\,({\rm at.ms})^{-1}$ to $10\,({\rm at.ms})^{-1}$, depending on the trapping and excitation  parameters explored. In order to compare our results to theoretical models and to measurements reported elsewhere in other trapping configurations, we calculate the normalized two-body loss rate $\beta =\beta'~2\sqrt{2}V$~\cite{WeinerRMP}, where $V=(\frac{2\pi k_B T_{0}}{m \omega^2})^{3/2}$ is the volume occupied by the atoms assumed to be at thermal equilibrium at $T_{0}$. Here, $\omega=(\omega_\perp^2\omega_\parallel)^{1/3}$ is the geometric average of the dipole trap oscillation frequencies $\omega_\perp$ and $\omega_\parallel$, and $m$ is the mass of an atom. For example, the data shown in Fig.~\ref{Fig:DecayCurve_Multi}a are best fitted when $\beta'= 1.2\pm 0.5$~(at.ms)$^{-1}$. Using $T_0=300~\mu$K and the parameters of our set-up ($\omega_\perp=130$~kHz and $\omega_\parallel=25$~kHz), we obtain $V=0.7~\mu$m$^3$ and $\beta=2.4\pm 1.1 \times 10^{-9}$~${\rm cm}^3.{\rm s}^{-1}$.

We extracted in the same way the loss rate $\beta$ for various values of the trap depth $U$ and of the excitation detuning $\delta$ and saturation $s$. Figure~\ref{Fig:SummaryBeta}a summarizes our results for a saturation parameter $s=0.5$ when we scan the frequency of the excitation light across the trap depth. We observe a resonance in $\beta$ that is shifted to the red with respect to the frequency corresponding to the bottom of the trap, i.e. when the light-shifted detuning $\Delta=\delta-U/\hbar\lesssim 0$. On the blue-side of the resonance ($\delta\gtrsim U/\hbar$), the two-body loss rate is suppressed, due to the excitation to a repulsive potential curve~\cite{WeinerRMP,Bali94}.

We also observe that the peak value of $\beta$ increases by more than an order of magnitude when the trap depth decreases only by a factor $3$. Figure~\ref{Fig:SummaryBeta}b shows that $\beta$ also increases as $s/(1+s)$, in qualitative agreement with a simple model assuming a two-level system. For the largest saturation parameter investigated ($s=1.5$), we find our largest value of the two-body rate constant $\beta=3.0\pm 1.5\times 10^{-8}$~${\rm cm}^3.{\rm s}^{-1}$.

\subsection{D. Discussion}
Our measurements indicate that the light-assisted two-body loss rate can reach values remarkably larger than any reported measurements we could find, using three-dimensional excitation light and either $^{85}$Rb or $^{87}$Rb. For example, Kuppens {\it et al.}~\cite{Kuppens2000} measure $\beta \sim 10^{-9}$~${\rm cm}^3.{\rm s}^{-1}$ using a dipole trap with a waist of $26~\mu$m and trap depths, temperatures and spatial densities comparable to ours. Kulatunga {\it et al.}~\cite{Kulatunga2010} use a dipole trap with a waist of $5.6~\mu$m size and measure two-body loss
rates as large as $\beta' \approx 10^{-2}\ ({\rm at.s})^{-1}$. Estimating their volume at thermal equilibrium, we have found that this corresponds to a normalized loss rate $\beta \sim 10^{-11}$~${\rm cm}^3.{\rm s}^{-1}$. Only Schlosser {\it et al.}~\cite{Schlosser2001,Schlosser2002} have to assume large values of $\beta' \sim 1000\ ({\rm at.s})^{-1}$ to explain the loading of at most one atom in their sub-micrometer size dipole trap. Again estimating their one-atom thermal volume~\cite{Tuchendler2008,Kuppens2000}, this yields $\beta \sim 3\times10^{-9}$~${\rm cm}^3.{\rm s}^{-1}$.

The analysis presented in Sec.~IV-C assumed the volume $V$ occupied by the atoms to be constant during the excitation pulse, and equal to the thermal volume for atoms at a constant temperature $T_0$. This assumption is actually not valid when the influence of the heating is larger than or comparable to the two-body loss mechanism, i.e. when $\delta\gtrsim U/\hbar$. However, neglecting the temperature increase during the light excitation leads to actually underestimate $V$ and thus $\beta$. For instance, we checked that for the highest values of $\beta$ that we measured the temperature increased by less than $15\%$ during the excitation, leading to an underestimation of $\beta$ by less than $25\%$, a difference within our error bars.

\subsection{E. Comparison to a semi-classical model}
Finally, we compare our largest measured light-assisted loss rate (i.e. $\beta_{\infty}=3.1\pm0.2\times10^{-8}$~${\rm cm}^3.{\rm s}^{-1}$ in Fig.~\ref{Fig:SummaryBeta}b) to the Langevin semi-classical limit $\beta_{\rm L}=\sigma_{\rm L} v$~\cite{JulienneVigue91} for the collision rate. Here, $v=\sqrt{\frac{16k_B T_0}{\pi m}}$ is the average velocity of the atoms in the frame of the two-body center of mass. The Langevin cross-section $\sigma_{\rm L}$ is obtained by summing the maximum cross-sections $(2l+1) 4\pi\hbar^2/(m E)$ up to the maximum partial wave $l_{\rm max}$ contributing to the collision, for a given collision energy $E$. In this approach,
\begin{equation}\label{eq:Langevin_cross_section}
    \sigma_{\rm L}=\frac{4\pi\hbar^2}{mE}(l_{\rm max}+1)^2.
\end{equation}
We calculate $l_{max}$ by imposing two conditions. First, assuming the pair of atoms has been excited in the $S+P$ potential (see Fig.~\ref{Fig:figure1}), the kinetic energy $E=\frac{3}{2}k_BT_0$ of the two colliding atoms in the frame of their center of mass must be larger than the height of the centrifugal barrier to allow the collision to take place at short interatomic distance. This condition yields
\begin{equation}\label{eq:lmax1}
    l_{\rm max,1}(l_{\rm max,1}+1)=\frac{3m}{2\hbar^2}(2C_3^2E)^{1/3},
\end{equation}
\begin{equation}\label{eq:sigma1}
    \sigma_{\rm L}\approx 6\pi\left(\frac{2C_3^2}{E^2}\right)^{1/3} \ ,
\end{equation}
as $l_{\rm max,1}\gg 1$ ($l_{\rm max,1}\simeq 50$ typically). Second, the height of the centrifugal barrier in the $S+S$ potential should be small enough to allow a pair of atoms with energy $E$ to be excited at an interatomic distance shorter than their minimal approach distance. This yields
\begin{equation}\label{eq:lmax2}
    l_{\rm max,2}(l_{\rm max,2}+1)\hbar^2/(m  r_\textrm{exc}^2)= E
\end{equation}
where the distance $r_\textrm{exc}$ actually depends on the light-shifted detuning $\Delta$ through $r_\textrm{exc}=\left(-C_3/\hbar\Delta\right)^{\frac{1}{3}}$. This second condition yields
\begin{equation}\label{eq:sigma2}
    \sigma_{\rm L} \approx 4\pi r_\textrm{exc}^2.
\end{equation}

The maximal partial wave contributing to the cross-section is actually $l_{\rm max}=\textrm{Min}(l_{\rm max,1},l_{\rm max,2}$). Figure~\ref{Fig:SummaryBeta} shows the Langevin limit set by these two conditions. For atoms at a temperature $T_0= 200~\mu $K in a trap with $U/h=18$~MHz and an excitation with $\delta/2\pi=10$~MHz, $\beta_{\rm L}= 4.1\times 10^{-8}$~${\rm cm}^3.{\rm s}^{-1}$, close to our largest measured value $\beta_{\infty}=3.1\pm0.2\times10^{-8}$~${\rm cm}^3.{\rm s}^{-1}$ (see Fig.~\ref{Fig:SummaryBeta}b). Given the simplicity of the model, it is quite surprising that the light-assisted process studied here approaches this theoretical limit: for alkalies the two-body collision rate is predicted to be smaller than the Langevin limit by at least one order of magnitude~\cite{JulienneVigue91}.

\section{V. Conclusion}
In conclusion, using our ability to isolate one-body radiative heating from two-body losses, we have measured remarkably large two-body collision rates in a micrometer size optical dipole trap in the presence of near resonant light. We have found that these large rates are close to the semi-classical Langevin limit.
Given the complexity of the situation considered here, due to the near-resonant character of the light combined to the small size of the trapping potential that may affect the interaction between the atoms, it is quite remarkable that a simple semi-classical argument reproduces our largest measured value. It would be interesting to cross-check our findings using atoms in optical lattices, a situation where the sites also have a sub-micrometer size and where the number of atoms per site can be controlled precisely~\cite{Bakr2010,Bloch2010}.

\begin{acknowledgments}
We acknowledge support from the E.U. through the ERC Starting Grant ARENA, and from IFRAF. A.~F. acknowledges partial support from the DAAD Doktorandenstipendium. We thank E.~Tiesinga, P.L.~Gould, J.~Vigué, B.~Laburthe-Tolra for discussions.
\end{acknowledgments}

\end{document}